\documentstyle[preprint,aps]{revtex}
\tighten
\begin{document}
\title{DEFORMATIONS  OF EXTENDED OBJECTS \\
 WITH EDGES}
\author{ Riccardo Capovilla${}^{1}$ and Jemal Guven${}^{2}$
\thanks{Permanent address:
Instituto de Ciencias Nucleares, 
 Universidad Nacional
Aut\'onoma de M\'exico, Apdo. Postal 70-543, 04510,
M\'exico, D.F., MEXICO}}
\address{ ${}^1$\ 
 Departamento de F\'{\i}sica,
 Centro de Investigacion
y de Estudios Avanzados del I.P.N. \\
Apdo Postal 14-740, 07000 M\'exico,
D. F., MEXICO \\
${}^2$\ 
School of Theoretical Physics, 
D.I.A.S. \\
10 Burlington Rd., Dublin 4, IRELAND 
}
\maketitle
\begin{abstract}
We present a manifestly gauge covariant description of fluctuations of 
a relativistic extended object described by the Dirac-Nambu-Goto action 
with Dirac-Nambu-Goto loaded edges about a given classical solution.
Whereas physical fluctuations of the bulk lie normal to its worldsheet, 
those on the edge possess an additional component directed into the bulk. 
These fluctuations couple in a non-trivial way involving the underlying 
geometrical structures associated with the worldsheet of the object 
and of its edge. We illustrate the formalism using as an example a 
string with massive point particles attached to its ends. 
\end{abstract}
\date{\today}
\vskip2pc

\section{Introduction}

Relativistic extended objects with physical edges
occur in many contexts. The simplest realization,
often exploited to model hadrons 
consists of  a string with particles attached to its ends (a zero
dimensional edge). 
This model has obvious higher dimensional 
analogues: a membrane bounded by a string (a one
dimensional edge) and so on. Once an edge is introduced
it will generally possess its own degrees of freedom and 
capable of exchanging energy with the bulk.
Indeed, to be democratic, these degrees of freedom should be 
treated on at least the same footing as those in the bulk.
Often one might wish to focus on the edge degrees of freedom,
integrating out the bulk degrees of freedom in favour of an action 
at a distance description of the edge dynamics. Such is the 
case in string models for QCD.
What is clear is that the dynamics will generally 
depend sensitively on the specific interaction 
we admit between the bulk and edge degrees of freedom.
The simplest action describing the dynamics of a relativistic extended 
object (membrane) with a non-null edge
is a sum of two Dirac-Nambu-Goto [DNG] terms, 
{\it i.e.} one proportional to the area of its
worldsheet, $m$, the other proportional to the area of 
the worldsheet, $\partial m$, of its boundary. The 
interaction induced between the bulk and the edge by this
action is local. While it is perhaps over optimistic to expect 
this model to describe realistic physical systems, 
it does provide a useful point of
departure for more sophisticated study.   

In earlier work, we demonstrated that the
interaction between the dynamical degrees of freedom on the edge
and those in the bulk can be cast in a 
universal geometrical form \cite{RMF,Edges}. Carter has also addressed the 
problem from a different point of view recognising that both the bulk
equations of motion and the boundary equations could be cast
in a single generalized `sail' equation \cite{C1}. Earlier relevant 
work in the context of string models for
QCD is summarized in Ref. \cite{Nest}. For the case
of hybrid topological defects in cosmology, see Ref. \cite{VS}.
The dynamics in the  interior is described by the 
system of non-linear hyperbolic partial differential
equations:

\begin{equation}
K^i =0\,.
\label{eq:k0}
\end{equation}
Here $K^i$ is the trace of the $i^{\rm th}$ extrinsic curvature of $m$ 
embedded in spacetime, one for each co-dimension of the embedding. 
These equations are unchanged with respect to those 
which correspond to extended objects with
empty boundary. The interior worldsheet is 
always extremal. It is the coupling at the boundary 
which is non-trivial. Each edge must satisfy

\begin{equation}
\mu_b k = - \mu \,,\label{eq:k1}
\end{equation}
where $k$ is the trace of the extrinsic 
curvature of the boundary worldsheet $\partial m$ embedded 
as a hypersurface in $m$. The tension in the membrane is $\mu$, that 
in its boundary is $\mu_b$  (if pointlike, this will be a mass).
Eq.(\ref{eq:k1}) is the only place where the tensions 
feature explicitly. The boundary worldsheet has constant 
mean extrinsic curvature.

The edge in turn informs the bulk how to 
move by constraining the projected extrinsic curvatures
of $m$ to vanish on $\partial m$,

\begin{equation}
{\cal H}^{ab} K^i_{ab} = 0\,,
\label{eq:k2}
\end{equation}
where ${\cal H}^{ab}$ denotes the projector
from $m$ onto $\partial m$.
In \cite{Edges}, we emphasized the different roles played by 
Eq.(\ref{eq:k1}) and Eqs. (\ref{eq:k2}) in mediating the 
interaction between the bulk and the edge degrees of freedom.

The system of equations  (\ref{eq:k0}), (\ref{eq:k1}) and 
(\ref{eq:k2}) is simultaneously worldsheet/boundary 
worldsheet diffeomorphism covariant. 
Unfortunately, however, the system is untractable unless
a large degree of symmetry is imposed. Even for 
a string bounded by point particles, the coupling 
to the particles spoils the linearity of the problem.
 
In this paper, we propose to examine the interaction 
between the bulk and the edge perturbatively: at face value, this simply 
involves linearizing the equations of motion, (\ref{eq:k0}) to
(\ref{eq:k2}), about a given classical solution. In practice,
however, it is not so straightforward.
The challenge here is to cast the linearized equations, 
like the equations of motion,
in a manifestly  covariant geometrical form.

We already know how to describe fluctuations in the bulk
for a DNG extended object
\cite{GUV,LF,CAR,CG1}. We know that physical fluctuations there
correspond to normal deformations of the bulk worldsheet. There is 
therefore one for each co-dimension associated with
the embedding of the worldsheet in spacetime. These
quantities satisfy a system of coupled hyperbolic 
partial differential equations with 
support on the worldsheet of the undeformed object.
The principal subtlety in the description adopted in
Ref. \cite{GUV} lies in the identification of the role of the twist 
potential of the background worldsheet in ensuring            
covariance under worldsheet normal rotations.

On the edge there is an additional possible mode 
of fluctuation directed into (or out of) the bulk, tangent
to the bulk worldsheet. These edge fluctuations will
couple to fluctuations in the bulk. 
They satisfy a hyperbolic partial (ordinary if pointlike) 
differential equation with support on the 
worldsheet of the undeformed edge and with a source 
linear in the boundary values of the bulk
modes. The boundary conditions on the bulk fluctuations are 
themselves dynamical, assuming the form of
a system of coupled hyperbolic partial 
differential equations with support on the 
worldsheet of the undeformed edge with source linear in 
the edge mode. Clearly, this 
is an intricately coupled system. Indeed, generically, it is
not possible to decouple them. Neither pure edge modes nor 
pure bulk modes with an independent dynamical life are 
supported by the system.

To derive the linearized equations of motion, our strategy is to first 
control technicalities of a kinematical
nature. For this reason, sect.II is devoted to
a derivation of some  kinematical identities in the manner
undertaken in Ref. \cite{CG1} for the parent worldsheet.
Specifically, we examine the deformations
in the edge geometry, both intrinsic and extrinsic, which are
induced by a deformation in the edge worldsheet.
There are two ways this deformation can be approached. The 
approach adopted here exploits the possibility of describing the 
edge  worldsheet itself  as an embedding  in spacetime by 
forming the composition of embeddings: embedding first edge worldsheet
in the bulk worldsheet followed by the 
embedding of the  bulk worldsheet in spacetime.
The alternative approach (the one adopted in Ref. \cite{Edges} for
calculations) focuses on the boundary behavior of the bulk worldsheet.
While this straightforward approach was adequate in that context,
where we needed only consider deformations of
intrinsic geometry, its shortcomings become 
apparent as soon as one moves on to treat deformations of the 
extrinsic geometry. Indeed the approach adopted here not only facilitates 
calculations, but it also further elucidates the geometrical nature 
of the coupling between  edge and bulk fluctuations.

We end with an application of our formalism. We first specialize to the 
case of a string bounded by massive point particles.
We then examine the special case of the motion of a rigidly rotating 
string with particles attached to its ends. 
We show that, contrary to naive Newtonian expectations,
which would suggest the existence of a breathing mode
corresponding to the motion of a particle in a linear potential,
the only such motion is circular corresponding to a 
timelike right circular helicoid. We note that the string 
equation of motion is not among the set of 
equations of motion corresponding to a truncation of the 
action to the symmetry appropriate to rigid motion.
We examine how this system behaves
under perturbations. 
We find that the normal modes of the 
perturbations have complex frequencies.

For the sake of simplicity, in the text we consider
only a flat background spacetime.
We relegate the equations corresponding
to the case of an arbitrary background spacetime
to an Appendix.

\section{ Embedding of the edge worldsheet in spacetime}

In  Ref. \cite{Edges}, we showed that, with respect to an
adapted basis of normal vectors, both the equations of 
motion (\ref{eq:k1}) and
the boundary conditions (\ref{eq:k2}) can be cast 
entirely in terms of the extrinsic 
geometry associated with the direct embedding
of the edge worldsheet in spacetime.  
This was treated as a curiosity in Ref. \cite{Edges}. 
Because of the central role this shift in 
perspective will play in the sequel, it 
is worthwhile to recall the treatment  
in Ref. \cite{Edges}, elaborating when appropriate. 

The embedding of the bulk worldsheet, $m$, in spacetime,
induces 
an embedding in spacetime of the edge worldsheet
as follows:  
we identify the edge worldsheet with the 
timelike boundary of $m$ ($\partial m$ say). This is  
described by the embedding
$\xi^a = \chi^a(u^A)$ ($a= 0,1,\cdots, D-1$, $A=0,1,\cdots,D-2$) in $m$. 
$m$ in turn is 
described by the embedding in spacetime, 
$x^\mu = X^\mu(\xi^a)$, ($\mu=0,1,\cdots,N-1$). The composition of 
these two embeddings yields an embedding of the edge worldsheet
in spacetime, 

\begin{equation}
x^\mu = \bar X^\mu(u^A) \,,
\label{eq:emb}
\end{equation}
where

\begin{equation}
 \bar X^\mu(u^A) = X^\mu (\xi^a (u^A))\,.
\label{eq:Xcomp}
\end{equation}
We denote the tangent vectors to $m$ in spacetime, $e^\mu_a$;
the tangent vectors to the edge worldsheet in $m$
associated with $\chi$, $\epsilon^a_A$. The 
latter can be 
promoted to  spacetime vectors associated with the embedding $\bar X$,

\begin{equation}
f^\mu{}_A := e^\mu{}_a \epsilon^a{}_A\,.
\label{eq:prom}
\end{equation}

The metric $h_{AB}$ induced directly from spacetime by the 
embedding $\bar X^\mu$ coincides with that induced from $m$ by the 
embedding $\chi^a$:

\begin{equation}
h_{AB} = g_{\mu\nu}f^\mu{}_A f^\nu{}_B = \gamma_{ab}
\epsilon^a{}_A\epsilon^b{}_B
\,.
\label{eq:hAB}
\end{equation}

To construct the extrinsic geometry associated 
with the embedding $\bar X^\mu$, we need to 
erect the normals to $\partial m$ in spacetime. 
The choice we make exploits the 
fact that the spacetime normals to $m$, $\{n^{\mu\,i}\}$
are also normal to $\partial m$ in spacetime
($ i,j, \cdots = 1,\cdots,N-D$). 
We take $\{n^{\mu\,i}\}$ to satisfy
$ g_{\mu\nu} n^{\mu\,i} n^{\nu\,j} = \delta^{ij}$. We now 
promote the unit normal to $\partial m$ into $m$ to a spacetime vector with
$\eta^\mu := f^\mu{}_a \eta^a$. 
The vector $\eta^\mu$ is clearly normal to 
the $\{n^{\mu\,i}\}$ in spacetime. We how have a complete 
orthonormal basis of normal 
vectors, which we label $m^I:= \{ m^0 , m^i \} = \{\eta, n^i\}$, such that 
$ g_{\mu\nu} m^{\mu\,I} m^{\nu\,J} = \delta^{IJ}$ ($ I,J, 
\cdots = 1,\cdots,N-D+1$). 
We use the index 0 to 
denote the direction along $\eta^\mu$. It should not
be confused with a timelike index. In adapting our choice of 
normals in this way, we surrender the full $O(N-D+1)$ gauge freedom 
associated with rotations of an arbitrarily 
chosen set of normal vectors, $\{m^{\mu\,I}\}$,
for the $O(N-D)$ rotation freedom possessed by the $\{n^{\mu\,i}\}$.

We now write down the 
Gauss-Weingarten equations associated with 
the embedding (\ref{eq:Xcomp}): 

\begin{eqnarray}
D_A f_B &=& \gamma_{AB}{}^C f_C - L_{AB}{}^I m_I\,,  
\label{eq:gauss2} \\
D_A m^I &=& L_{AB}{}^I f^B + \sigma_{A}{}^{IJ} m_J\,.
\label{eq:wein2}
\end{eqnarray}
Here $D_A:=f^\mu{}_A D_\mu$, with
$D_\mu$  the spacetime covariant derivative. 
The connection $\gamma_{AB}{}^C$ is the one
compatible with $h_{AB}$. The quantity $L_{AB}{}^I
= L_{BA}{}^I$ is the $I^{\rm th}$
extrinsic curvature of $\partial m$ embedded in spacetime,
and $\sigma_A{}^{IJ}$ its extrinsic twist potential.

With respect to the adapted basis, it is simple to check that

\begin{equation}
L^i_{AB}= \epsilon^a_A\epsilon^b_B K^i_{ab} =K_{AB}{}^i\,,
\label{eq:KAB}
\end{equation}
and that

\begin{equation}
\sigma_{A}{}^{ij} = \epsilon^a_A \omega_{a}{}^{ij}
= \omega_A{}^{ij}\,.
\end{equation}
The boundary inherits the 
extrinsic curvature $K_{ab}{}^i$ and extrinsic twist 
$\omega_a{}^{ij}$ associated with
the embedding of the parent worldsheet $m$ in spacetime.

The component of the extrinsic curvature
directed along the normal $\eta^\mu$
is the extrinsic curvature associated with the 
embedding of the edge in the bulk:

\begin{equation}
L_{AB}{}^0 = k_{AB}\,..
\end{equation}
Thus 

\begin{equation}
k = h^{AB} k_{AB} = h^{AB} L_{AB}{}^0\,.
\end{equation} 
Recalling that ${\cal H}^{ab} = h^{AB}\epsilon^a{}_A \epsilon^b{}_B $ is the
projector from the bulk onto the edge,
we also have  

\begin{equation}
{\cal H}^{ab} K_{ab}{}^i = h^{AB} L_{AB}{}^i\,.
\end{equation}
It is now clear that Eqs.(\ref{eq:k1}) and 
(\ref{eq:k2}) can be recast as

\begin{equation}
\mu_b L^I = - \mu \delta^{I0} \,.
\label{eq:LI}
\end{equation}

To complete the description of the boundary twist, we require
$\sigma_{A}{}^{i0}$. This is constructed from the tangent-normal
projection of the bulk worldsheet extrinsic curvature
as

\begin{equation}
\sigma_{A}{}^{i0} =\eta^a \epsilon^b_A K_{ab}{}^i =: K_A{}^i\,.
\label{eq:omega}
\end{equation}
Thus, part of the bulk 
worldsheet extrinsic curvature gets cast up on the boundary as a 
component of the twist potential. 
Henceforth we will denote this projection with $K_A{}^i$. 
The fact that $\sigma_{A}{}^{i0}$ is completely determined 
is consistent with the fact that 
we have surrendered the freedom to rotate $\eta^\mu$ into the 
$n^\mu_i$'s. Adapting the basis amounts to a partial gauge choice.
We note that the only projections of the bulk 
worldsheet extrinsic curvature which have not
been picked up by the boundary extrinsic geometry 
are the $K^i_{ab}\eta^a\eta^b$. For a 
DNG-DNG system this projection vanishes on the 
boundary and therefore the entire parent worldsheet extrinsic 
geometry in the neighborhood of the boundary is 
encoded completely in the boundary twist.

\section{Deforming the edge geometry}

Let us consider a deformation of the edge worldsheet,
$\partial m$, 
described by the infinitesimal deformation in the embedding
(\ref{eq:emb}),

\begin{equation}
\bar X^\mu \to \bar X^\mu + \delta \bar X^\mu\,.
\label{eq:defbd}
\end{equation}
We can decompose it with respect to the spacetime basis
adapted to the edge, 

\begin{equation}
\delta \bar X = \phi_I m^I + \phi^A f_A\,,
\end{equation}
{\it i.e.} in parts normal and tangential
to $\partial m$, respectively.
The latter term corresponds to a boundary worldsheet 
diffeomorphism which we subsequently drop, since we will
be interested only in deformations of quantities invariant
under reparameterizations of the edge worldsheet.
 The normal
deformation decomposes naturally into two parts:
one part normal to the parent worldsheet, the other
directed into it. We write: 

\begin{equation}
\phi^I m_I = \psi \eta + \phi^i n_i\,.
\label{eq:defb}
\end{equation}
$\phi^i$ can be identified with the value assumed on the boundary 
of the projection along the worldsheet 
normal $n_i$ of a deformation occuring in the bulk:

\begin{eqnarray}
X^\mu &\to&  X^\mu + \delta  X^\mu\,, \nonumber \\
\delta X &=& \Phi^i n_i + \Phi^a e_a\,, \nonumber \\
\phi^a &=& \Phi^a |_{\partial m} = \phi^A \epsilon^a{}_A
+ \psi \eta^a\,, \nonumber \\
\phi^i &=& \Phi^i |_{\partial m}\,.
\end{eqnarray}
We will subsequently refer to $\Phi^i$ as the bulk normal deformation.
>From the point of view of the parent worldsheet, $\psi$
corresponds to a piece of its tangential deformation;
 we will subsequently refer to it as the edge normal deformation.
Note that we are not allowed to drop the tangential part of
the deformation of the bulk worldsheet, since the presence
of a non-empty boundary breaks diffeomorphism invariance.
This is perhaps best appreciated if one interprets 
diffeomorphisms  as active transformations which can push 
a point  off the edges. 
 
We can now exploit the general formalism 
developed in \cite{CG1} to describe 
deformations of an arbitrary worldsheet, and apply it
to the deformation of the edge geometry. 

To simplify our treatment, in the text we will consider
only
the case of a flat background spacetime. The modified
expressions one obtains in the case of an arbitrary
background spacetime are given in an Appendix.

We have that, under the deformation (\ref{eq:defb}) the metric 
varies according to

\begin{equation}
D_{\delta\bar X_\perp} h_{AB} =  2 L_{AB}{}^I \phi_I 
= 2  K_{AB}{}^i \phi_i + 2 k_{AB}\psi\,,
\label{eq:delh}
\end{equation}
where we use the deformation operator

\begin{equation}
D_{\delta\bar X_\perp} := (\delta \bar X^\mu{})_\perp D_\mu =
\phi^I m^\mu{}_I D_\mu = \psi \eta^\mu D_\mu + \phi^i n^\mu{}_i D_\mu\,.
\end{equation}

The extrinsic curvature of the edge as embedded in spacetime 
transforms as follows\cite{CG1}: 

\begin{equation}
\hat{D}_{\delta\bar X_\perp} L_{AB}{}^I
= -\hat{\cal D}_A \hat{\cal D}_B \phi^I +  L_{AC}{}^I 
L^C{}_{B\,J}  \phi^J\,.
\label{eq:delL}
\end{equation}
On the right hand side we use the covariant derivative associated with the
connection $\sigma_A{}^{IJ}$,

\begin{equation}
\hat{\cal D}_A  \phi^I = 
{\cal D}_A  \phi^I - \sigma_A{}^{IJ} \phi_J\,,
\end{equation}
where ${\cal D}_A$ is the covariant derivative on
the edge, compatible with $h_{AB}$.
The hat appearing over the deformation operator
 $D_{\bar X_\perp}$ is the 
analogue for normal deformations of $\bar X^\mu$ of the 
totally covariant operator 
$\tilde{\cal D}_\delta$ for normal
deformations of $X^\mu$ which was introduced in Ref. \cite{CG1}.
Its significance will be explained below.

Our strategy is to
first decompose Eq.(\ref{eq:delL})
into mutually orthogonal parts 
with respect to the decomposition (\ref{eq:defb}). 
We express the $I=0$ component of Eq.(\ref{eq:delL})
as

\begin{equation}
\hat D_{\delta\bar X_\perp} L_{AB}{}^0 = 
\hat D_{\delta\bar X_\perp} k_{AB}
= - \hat{\cal D}_A\hat{\cal D}_B \psi +
 k_{AC} k^C{}_{B}  \psi 
+  k_{AC}  K^C_{B\,i} \phi^i
\,.
\label{eq:delk}
\end{equation}
We now exploit the fact that

\begin{equation}
\hat{\cal D}_A \phi^i =
\widetilde{\cal D}_A \phi^i - K_A{}^i \psi\,,
\label{eq:p1}
\end{equation}
where we have used the notation

\begin{equation}
\widetilde{\cal D}_A \phi^i := \epsilon^a{}_A \widetilde\nabla_a \phi^i 
= \epsilon^a{}_A \left[ \nabla_a \phi^i - 
\omega_a{}^{ij}\phi_j \right]\,,
\end{equation}
and that

\begin{equation}
\hat{\cal D}_A \psi = {\cal D}_A \psi + K^i_A \phi_i\,.
\label{eq:p2}
\end{equation}
With the help of these expressions, 
 we can decompose the first term appearing in Eq.(\ref{eq:delk}),

\begin{eqnarray}
\hat{\cal D}_A \hat{\cal D}_B \psi 
&=&
{\cal D}_A ({\cal D}_B \psi + K^i_B \phi_i) + K^i_A
(\widetilde{\cal D}_B \phi_i - K_{i\,B}\psi)\nonumber\\
&=& 
{\cal D}_A {\cal D}_B \psi + 2 K^i_{(A}\widetilde{\cal D}_{B)}\phi_i
+ ( \widetilde{\cal D}_A K^i_B ) \, \phi_i 
- K^i_A K_{i\,B}\psi\,,
\label{eq:hess}
\end{eqnarray}
so that Eq. (\ref{eq:delk}) can now be written in the form

\begin{eqnarray}
\hat D_{\delta\bar X_\perp} k_{AB}
=   &-& {\cal D}_A {\cal D}_B \psi + \left[ K^i_A K_{i\,B} 
+ k_{AC} k^C{}_B \right]  \psi \nonumber \\
&-& 2 K^i{}_{(A}\widetilde{\cal D}_{B)}\phi_i
- (\widetilde{\cal D}_A K^i_B ) \, \phi_i 
+  k_{AC}  K^C_{B\,i} \phi^i\,.
\label{eq:delkab}
\end{eqnarray}

We can exploit Eq.(\ref{eq:delh}), together with 
Eq.(\ref{eq:delkab}) to obtain for the 
deformation of the trace,

\begin{eqnarray}
\hat{D}_{\delta\bar X_\perp} k = 
 &-& {\cal D}_A{\cal D}^A \psi 
+ \left[K_{A\,i} K^{A\,i} - k_{AB}k^{AB}  
\right]\psi\nonumber\\
&-&  
2 K_{A}{}^i (\widetilde{\cal D}^A\phi_i )
- (\widetilde{\cal D}_A K^{A\,i} ) \, \phi_i
- k_{AB} K^{AB}{}_i  \phi^i \,.
\label{eq:deltrk}
\end{eqnarray}

Let us now consider the transformation of the remaining 
projections, $L_{AB}{}^i$.
>From Eq. (\ref{eq:delL}), we have that

\begin{equation}
\hat{D}_{\bar X_\perp} L_{AB}{}^i = 
\hat{D}_{\bar X_\perp} K_{AB}{}^i
= -\hat{\cal D}_A \hat{\cal D}_B \phi^i +
 K_{AC}{}^i k^C{}_B \psi + K_{AC}{}^i K^C{}_{B\,j} \phi^j
\,.
\label{eq:dlabi}
\end{equation}
Here, using Eqs. (\ref{eq:p1}), (\ref{eq:p2}), we can express the first term
on the right hand side as

\begin{eqnarray}
\hat{\cal D}_A \hat{\cal D}_B \, \phi^i 
&=&
\widetilde{\cal D}_A (\widetilde{\cal D}_B \phi^i
 - K_B{}^i\, \psi)
- K_A{}^i ({\cal D}_B \psi - K_B{}^j \, \phi_j)
\nonumber\\
&=& 
\widetilde{\cal D}_A\widetilde{\cal D}_B \phi^i 
- 2 K_{(A}{}^i {\cal D}_{B)}\psi
- (\widetilde{\cal D}_A K_B{}^i ) \, \psi 
-K_A{}^i K_{B\,j} \phi^j\,,
\end{eqnarray}
so that we find 

\begin{eqnarray}
\hat{D}_{\delta\bar X_\perp} K_{AB}{}^i
= &-&\widetilde{\cal D}_A \widetilde{\cal D}_B \phi^i +
\left[ K_{AC}{}^i K^C{}_{B\,j} + K_A{}^i K_{B\,j}\phi^j \right]
\nonumber\\
&+&  K_{AC}{}^i k^C{}_{B} \psi + (\widetilde{\cal D}_A K_B{}^i) \psi +
2 K_{(A}{}^i {\cal D}_{B)}\psi\,.
\label{delKAB}
\end{eqnarray}

Finally, for the deformation of its trace over the edge indices
we obtain 

\begin{eqnarray}
\hat{D}_{\delta\bar X_\perp} \left( h^{AB} K_{AB}{}^i \right)
= &-&\widetilde{\cal D}_A \widetilde{\cal D}^A \phi^i -
\left[ K_{AB}{}^i K^{AB\,j} -  K_A{}^i K^A{}_j \right] \phi^j \nonumber\\
&-&  K_{AB}{}^i k^{AB}\psi  - (\widetilde{\cal D}_A K^{A\,i} ) \psi 
+   2 K_{A}{}^i {\cal D}^A\psi\,.
\label{dtrKAB}
\end{eqnarray}

This is all we need for DNG extended objects with
DNG edges. It is straightforward to use this method
to obtain the expressions for the deformation of
the remaining geometrical
structures on the boundary.

Let us return now to the promised
explanation of the significance of the
hat on the deformation operator  $D_{\bar X_\perp}$.
It is related to  
its covariance transformation properties  under
normal rotations. 

First, we consider the following calculational
check on  Eq. (\ref{eq:delkab}). If we let $\phi^i = 0$ in
Eq. (\ref{eq:delkab}), that is if we specialize to pure
normal deformations of the edge extrinsic 
curvature (into $m$), we should
get agreement with  the normal deformation of the
extrinsic curvature of $\partial m$, seen as an
hypersurface in $m$.
By specializing Eq.(4.6) of 
Ref. \cite{CG1} to the case of an hypersurface, this is 
given by 

\begin{equation}
D_{\delta\chi_\perp} k_{AB}
= - {\cal D}_A{\cal D}_B \psi 
+ \left[k_{AC}k^C{}_B - 
{\cal R}_{acbd} \,\epsilon^a_A\epsilon^b_B  
\eta^c\eta^d\right]\psi\,,
\label{eq:dbkab}
\end{equation}
where ${\cal R}_{abcd}$ denotes the Riemann curvature
of $m$.
There is no twist and therefore no ambiguity in the 
definition of $D_{\delta\chi_\perp} := \psi \eta^\mu D_\mu$. 

To compare, we exploit the 
bulk Gauss-Codazzi equation
to express the 
worldsheet Riemann tensor projection
appearing in Eq.(\ref{eq:dbkab})
in terms of the appropriate projections
of worldsheet 
extrinsic curvature quadratics:

\begin{equation}
{\cal R}_{acbd} \,\epsilon^a_A\epsilon^b_B  \eta^c\eta^d
- K_{AB}{}^i K_{cd\,i} \eta^c \eta^d + K_{A}{}^i K_{B\,i} = 0\,.
\end{equation}
We note that the middle term vanishes for a 
DNG solution modulo the bulk equations
of motion (\ref{eq:k0}),  together 
with the boundary conditions (\ref{eq:k2}).
The last term coincides with an equal term 
appearing in the decomposition of the Hessian (\ref{eq:hess}).
We now have, using this expression to eliminate
the worldsheet Riemann curvature, that 

\begin{equation}
D_{\delta\chi_\perp} k_{AB}
= - {\cal D}_A{\cal D}_B \psi 
+ \left[k_{AC}k^C{}_B 
 + K_{A}{}^i K_{B\,i} - K_{AB}{}^i K_{cd\,i} \eta^c \eta^d \right]\psi\,.
\label{eq:dbkab1}
\end{equation}
Comparing now (\ref{eq:dbkab}) when $\phi^i = 0$,
 with (\ref{eq:dbkab1}),  the existence of the 
last term in the right hand side of Eq. (\ref{eq:dbkab1}) would appear
to point to an inconsistency. It is 
clearly not legitimate at this level 
to invoke the bulk equations of motion
to disregard it.  
 
In fact, there is no inconsistency. Here is where
the hat appearing on the deformation
operator  $D_{\delta\bar X_\perp}$ comes in. Recall 
that for a field transforming as a tensor under normal rotations,
such as $L_{AB}{}^I$, the construction of a totally covariant 
measure of the deformation 
$\hat{D}_{\bar X_\perp}$ involves 
the  addition of a deformation connection \cite{CG1}:

\begin{equation}
\hat D_{\delta \bar X_\perp} L_{AB}{}^I =
D_{\delta \bar X_\perp} L_{AB}{}^I 
- \hat\gamma^{IJ} L_{AB\,J}\,.
\end{equation} 
The connection is given by

\begin{equation}
\hat\gamma^{IJ} = g(D_{\delta \bar X_\perp} m^I, m^J)\,.
\end{equation}
This connection is necessary to ensure manifest covariance 
under the full $O(N-D+1)$ normal rotations.  
However, as emphasized above, we are breaking the 
invariance down to $O(N-D)$, when we surrender the freedom
to rotate the $\eta^\mu$ into the $n^{\mu\,i}$. 
This fact is responsible for the appearance of the
extra term in Eq. (\ref{eq:dbkab1}).

Let us, now, verify  explicitly
that the additional term appearing in 
Eq.(\ref{eq:dbkab1}) indeed coincides with the deformed 
connection correction. We have

\begin{equation}
\hat D_{\bar X_\perp} k_{AB}
= D_{\bar X_\perp} k_{AB}
- \hat\gamma^{0i} K_{AB\,i}\,.
\end{equation}
Using the ``deformation Gauss-Weingarten equations"
\cite{CG1}, a short calculation gives  

\begin{equation}
\hat\gamma^{0i} = - K_{ab\,i}\eta^a\eta^b \psi + 
 \eta\cdot\tilde\nabla\phi_i \,.
\label{eq:goi}
\end{equation}

The first term coincides with the offending term appearing in 
Eq.(\ref{eq:dbkab1}). The latter term involving normal 
gradients of the $\phi^i$ is of a form which does not appear 
elsewhere in Eq.(\ref{eq:delkab}), but that vanishes
anyway
when we consider pure normal deformations of the edge 
into $m$.

It was argued incorrectly in
Ref. \cite{CG1} that the deformation connection could 
be gauged away in general. 
We are presenting here an explicit counterexample to that
claim.
The argument outlining the 
circumstances under which it could be ignored 
was, however, correct. We always envisaged the 
application of the formalism to  
quantities which vanish on shell, such as,
for example, $K^i$. Also here, on shell, the deformation
connection does not contribute for example to the
deformation of the mean extrinsic curvature of the
edges, $k$, that is
$ \hat{D}_{\delta\bar X_\perp} k = D_{\delta\bar X_\perp} k$.

Similarly, for the deformation of the remaining projections,
$L_{AB}{}^i$, one has that

\begin{equation}
\hat{D}_{\delta\bar X_\perp} K_{AB}^i
= D_{\delta\bar X_\perp} K_{AB}^i - 
\hat \gamma^{ij} K_{AB\,j}
- \hat \gamma^{i0} k_{AB}\,,
\end{equation}

Noting that

\begin{equation}
\hat{\gamma}^{ij} = \gamma^{ij} 
+ \psi \eta^a \omega_a{}^{ij}, 
\end{equation}
and using Eq. (\ref{eq:goi}), one obtains

\begin{equation}
\hat{D}_{\delta\bar X_\perp} K_{AB}{}^i
= \widetilde{D}_{\delta\bar X_\perp} K_{AB}{}^i
+ [(\eta\cdot\tilde\nabla)\phi^i + 
K_{ab}{}^i\eta^a\eta^b 
] k_{AB} + \psi \omega^{ij}_a \eta^a)
K_{AB\,j}\,.
\end{equation}

For the deformation of its trace over the edge
indices, we obtain then

\begin{equation}
\hat{D}_{\delta\bar X_\perp}(h^{AB} K_{AB}{}^i )
= \widetilde{D}_{\delta\bar X_\perp} (h^{AB} K_{AB}{}^i )
+ [(\eta\cdot\tilde\nabla)\phi^i + 
K_{ab}{}^i\eta^a\eta^b ] k\,.
\end{equation}

\section{Linearized equations of motion}

It is straightforward at this point to read off 
the linearized equations of  motion. 
We have the linearized equation in the bulk, or
$\widetilde{D}_{\delta X_\perp} K^i = 0$, given by
\cite{GUV,LF,CAR,CG1}

\begin{equation}
\tilde\Delta \Phi^i + 
 K_{ab}{}^i K^{ab}{}_j  \Phi^j = 0 \,.
\label{eq:madre}
\end{equation}

The linearizations of Eq.(\ref{eq:k1}) and 
Eq.(\ref{eq:k2}) are respectively,
$\hat{D}_{\delta\bar{X}_\perp} k=0$, and 
$\hat{D}_{\delta\bar{X}_\perp} h^{AB} K^i_{AB} = 0$, or

\begin{eqnarray}
{\cal D}^A \left({\cal D}_A \psi + K^i_A\phi_i\right)
+ K^A_i\left(\widetilde{\cal D}_A\phi^i - K^i_A\psi\right) 
+ k^{AB} (k_{AB} \psi + K_{AB}^i \phi_i ) 
&=&0 \,, 
\label{eq:motion2} \\
\widetilde{\cal D}^A  \left(
\widetilde{\cal D}_A \phi^i - K_A^i \psi\right)  
- K^{A\,i}\left({\cal D}_A \psi + K^j_A\phi_j\right) 
+ K^i_{AB} (K^{AB\,j}\phi_j + k^{AB} \psi) &=& 0
\,.
\label{eq:motion3}
\end{eqnarray}
In this form, a symmetry in the boundary conditions
becomes manifest which is not apparent in their 
original form.

If the parent worldsheet is totally geodesic, so that
$K^i_{ab}=0$, then these equations reduce to

\begin{equation}
\tilde\Delta \Phi^i = 0 \,,
\label{eq:mama0}
\end{equation}

\begin{equation}
\left({\cal D}^A{\cal D}_A + k^{AB} k_{AB} 
\right)\psi =0\,,
\label{eq:psi}
\end{equation}

\begin{equation}
\widetilde{\cal D}^A
\widetilde{\cal D}_A \phi^i  
=0\,.
\end{equation}
Then $\phi^i$ and $\psi$ completely decouple. 
An example is provided
by a non - rotating straight length of string bounded by particles,
or a flat disc of membrane bounded by a string. 
In the former case, we have 
$\ddot\psi - (\mu/M)^2\psi =0$ and $\ddot \phi^i =0$ on the end, where
dot represents the derivative with respect to proper time
along the trabectory of the end particle of mass $M$ 
(also see next section).
The  tension in the bulk accelerates the end particles uniformly, it is 
necessarily destabilizing. We will discuss this issue 
in the next section.

Does any analogous behavior prevail when $K_{ab}^i \ne 0$?
Suppose we have $\Phi^i =  0 = \phi^i$ everywhere. Then 
the linearized equations of motion reduce to Eq. 
(\ref{eq:psi}), and

\begin{eqnarray}
\left[ {\cal D}^A {\cal D}_A   
+ k^{AB} k_{AB} - K^{Ai} K_{Ai} \right] \psi &=& 0\,,
\label{eq:psi1} \\ 
 2 K^{A\,i} {\cal D}_A \psi 
- ( \widetilde{\cal D}_A  K_A^i ) \psi   
+ K^i_{AB} k^{AB} \psi  &=& 0
\,.
\label{eq:psi2}
\end{eqnarray}

In general, the only solution of 
Eq.(\ref{eq:psi2}) which is consistent with the  Eq.(\ref{eq:psi1}) 
is the trivial solution, $\psi=0$.
Pure edge states do not generally exist.  
Conversely, pure bulk states with
$\psi =0$ are also impossible.

Of course, there remains the possibility to
fix the ends with $\psi = \phi^i = 0 $, and to
consider only bulk deformations, given by solutions
of (\ref{eq:madre}) with Dirichlet-type boundary
conditions at the ends.

\section{String bounded by point particles}

In this section, we specialize the general treament
given above, to the case
of a string with massive point particles attached to its
ends. 

We parametrize the boundary worldline by proper time, $\tau$.
We denote the corresponding unit velocity vector with $v^a$.
On the boundary, 
$\partial m$, the string worldsheet metric can be parametrized 
as

\begin{equation}
\gamma^{ab} = -v^a v^b + \eta^a \eta^b\,.
\end{equation}
where recall that $\eta^a$ is the normal to
$\partial m$ into $m$.

We define 

\begin{eqnarray}
K_{\parallel\parallel}{}^i &:=& K^i_{ab}v^a v^b\,, \\
K_{\perp\parallel}{}^i  &:=&  K_{ab}^i \eta^a v^b\,, \\
K_{\perp\perp}{}^i &:=& K^i_{ab}\eta^a\eta^b\,. 
\end{eqnarray}
The string background equations of motion,
evaluated at the boundary, then imply

\begin{equation}
K_{\perp\perp}{}^i = K_{\parallel\parallel}{}^i\,,
\end{equation} 
whereas the boundary conditions,
Eq.(\ref{eq:k2}), imply that

\begin{equation}
K_{\parallel\parallel}{}^i  = 0\,,
\label{eq:kpp}
\end{equation}
and as a result also 

\begin{equation}
K_{\perp\perp}{}^i=0\,.
\end{equation}
Thus, we can now 
express the 
extrinsic curvature of the worldsheet $m$ on the boundary in the 
off-diagonal form,

\begin{equation}
K_{ab}{}^i = - K_{\perp\parallel}{}^i  (\eta_a v_b + \eta_b v_a)\,.
\label{eq:strK}
\end{equation}
We then have on the boundary

\begin{equation}
K_{ab}{}^i K^{ab}{}_j = - 2 K_{\perp\parallel}{}^i
K_{\perp\parallel \, j} \le 0  \,.
\label{eq:str}
\end{equation}
The latter implies that 
the mass matrix appearing in the 
linearized bulk equations of equation, Eq. 
(\ref{eq:madre}), has no negative eigenvalues,
corresponding to tachyonic modes, in the neighborhood of the boundary.
 
We also note that the extrinsic curvature associated with the 
embedding of the end worldline in the string worldsheet is
identified with the geodesic curvature which is completely 
determined by the boundary conditions. We do not need to solve 
Eq.(\ref{eq:k1}) explicitly.
The boundary conditions, Eqs.(\ref{eq:motion2}) and (\ref{eq:motion3}), 
reduce to the system of coupled ordinary differential equations:

\begin{eqnarray}
\ddot \psi - \left[ K_{\perp\parallel}^i 
K_{\perp\parallel\,i} + k^2 \right]  \psi 
+ 2 K_{\perp\parallel}{}^i \dot{\phi}_i +
\dot K_{\perp\parallel}{}^i \phi_i 
- \omega_{\parallel\, ij}K_{\perp\parallel}{}^i \phi^j  &=& 0\,, \\
\widetilde{\ddot\phi}{}^i - K_{\perp\parallel}{}^i 
K_{\perp\parallel}{}^j \phi_j 
- 2 K_{\perp\parallel}{}^i \dot\psi - \dot{K}_{\perp||}{}^i \psi  
+ \omega_{\parallel\, ij} K_{\perp\parallel}{}^i \phi^j &=& 0\,,
\end{eqnarray}
where $\omega_{\parallel\, ij} := \eta^a \omega_{a\, ij}$.
The mass terms appearing in the boundary conditions 
should be contrasted with those in the bulk 
equations. Their diagonal entries
are explicitly tachyonic. This does not, in itself,
however, signal the non-existence of harmonic
perturbations. One has to take
into account 
both the off diagonal terms (coupling 
$\psi$ and $\phi^i$) and the first derivative terms.

\section{Rigidly rotating string bounded by point particles}

Consider a string bounded by two
pointlike masses rotating rigidly in a plane
\cite{Nest}. 
We describe Minkowski space by cylindrical polar coordinates,
$(t, r, \theta,z)$. The worldsheet 
generated by the motion of the string can be described by the embedding
in Minkowski space,

\begin{equation}
X^\mu ( t , r ) = 
\left(
 t\,, r \cos \theta (t) \,, r \sin \theta (t) \,,0 \right)\,.
\label{eq:emb1}
\end{equation}
This embedding induces a line element on the string 
worldsheet given by

\begin{equation}
ds^2 = - (1-r^2\dot\theta^2) dt^2 + dr^2 \,,
\label{eq:line1}
\end{equation}
for $0\le r\le R(t)$.

We now evaluate the extrinsic curvature tensor
which corresponds to the worldsheet described by 
Eq.(\ref{eq:line1}). It is clear that the only non-vanishing 
normal component is that which corresponds to the normal which lies in the 
plane of motion. We have 

\begin{eqnarray}
K_{tt} &= & {- r \ddot \theta\over (1-r^2\dot\theta^2)^{1/2}}
\nonumber\\
K_{rt} &=& { - \dot\theta \over (1-r^2\dot\theta^2)^{1/2}}\nonumber\\
K_{rr} &=& 0\,.
\end{eqnarray}
The extrinsic twist for this embedding is identically zero,
$\omega_a{}^{ij} = 0$.

The string equation of motion reduces to
$K_{tt}=0$ or $\ddot\theta =0$ with solution
$\theta = \omega_0 t + \theta_0$. This corresponds to a right circular
timelike helicoid, which is null when 
$\omega_0 r = 1$.

The proper time along the boundary of the worldsheet at $r= R(t)$ is 
in turn given by

\begin{equation}
d\tau^2 =  (1- \dot R^2 - R^2 \omega_0^2) dt^2 \,.
\label{eq:line2}
\end{equation}

 The tangent vector to the 
boundary at $r=R(t)$ is given by

\begin{equation}
v^a ( \tau ) = {1 \over \sqrt{1-\omega_0^2 R^2 + \dot R^2}}
\left(
 1 \,,  \dot{R} \right)\,,
\label{eq:va}
\end{equation}
while the normal into $m$ of the boundary is

\begin{equation}
\eta^a ( \tau ) = {1 \over \sqrt{1 - \omega_0^2 R^2} 
\sqrt{1-\omega_0^2 R^2 + \dot R^2}}
\left(
 \dot{R} \,,
 1 - R^2 \omega_0^2 \right)\,,
\label{eq:eta}
\end{equation}

The  projections of the parent worldsheet
extrinsic curvature onto the boundary is

\begin{equation}
K_{\parallel\parallel} = - {1 \over  (1 - \dot{R}^2
- R^2 \omega_0^2 ) \sqrt{1 - R^2 \omega_0^2}} 
2\omega_0 \dot{R}\,, 
\end{equation}
so that the boundary condition
implies that $R$ is constant (see Eq. (\ref{eq:kpp})).

We conclude that the equations of motion place a very stringent 
restriction on the possible motion. In particular,
Newtonian intuition is misleading.
It would suggest that rigid motion is possible
with variable $\dot\theta$ and variable $R$. 
After all the particles
at the ends themselves are subject to a linear potential, the 
rotation supplying a centrifugal repulsion so that
we have a straighforward Kepler problem. 
Indeed, one can formulate a Nambu theory 
for this system with a richer configuration space.
It is obtained by introducing (\ref{eq:line1}) and 
(\ref{eq:line2}) directly at the
level of the action \cite{ACG}.

Modulo the boundary conditions, one finds that $
K_{\perp\perp} = 0 $, as was to be expected,
and that the mixed projection is given by

\begin{equation}
K_{\perp\parallel} = {- \omega_0 \over
(1 - R^2 \omega_0^2 )^2 }\,.
\label{eq:kmx}
\end{equation} 

Finally, using that $R$ and $\dot\theta$ are constants,
the geodesic curvature of the ends is,

\begin{equation}
k = - { R \omega_0^2 \over 1 - R^2 \omega_0^2 }
\label{eq:gc}
\end{equation}
so that the ends equation of motion is

\begin{equation}
M R \omega_0^2 = \mu ( 1 - R^2 \omega_0^2 )\,.
\end{equation}

Let us now examine arbitrary small 
perturbations about such timelike circular helicoids. 
We have that

\begin{equation}
K_{ab}^i K^{ab}_j =
- 2\delta^i_1 \delta_j^1 { \omega_0^2 \over (1 - \omega_0^2 r^2)^{2}}\,.
\end{equation}
The mass matrix appearing in (\ref{eq:madre}) has one positive 
eigenvalue. Note that in the limit, $r \to R$, and
$\omega_0 R \to 1$, this curvature invariant diverges. The worldsheet
is singular in the neighborhood of its null boundary. It is
not surprising: there are no curvature 
penalties in the DNG dynamics. The DNG approximation breaks down.

We can now exploit a conformal coordinate system on the 
helicoid to simplify the form of the d'Alembertian
appearing in (\ref{eq:madre}). 
We have 

\begin{eqnarray}
ds^2  &=& (1 -\omega_0^2 r^2)  (- dt^2 + {dr^2 \over 1- 
\omega_0^2 r^2} )
\nonumber\\
      &=&  \cos^2 (\omega_0 X) (-dt^2 + dX^2)\,,
\end{eqnarray}
where

\begin{equation}
\omega_0 X =  \arcsin ({\omega_0 r})\,.
\end{equation}
Thus, the bulk linearized equations of motion
reduce to 

\begin{eqnarray}
{\partial^2 \Phi_1  \over \partial t^2 }
- {\partial^2 \Phi_1 \over \partial X^2} 
+ {2\omega_0^2\over \cos ^2
 (\omega_0 X)}\, \Phi_1 &=& 0\,, \\
{\partial^2 \Phi_2  \over \partial t^2 }
- {\partial^2 \Phi_2 \over \partial X^2} &=& 0\,, 
\label{eq:M1}
\end{eqnarray}
Note that they are completely decoupled, and that the
solutions are well behaved. On a solution without dynamical 
boundary, the solution appears to be stable. However, 
the geometry is singular at the null edge and it 
is not obvious that perturbation theory makes sense.

On the boundary, we have, with $\phi := \phi_1$,

\begin{eqnarray}
\ddot \psi - (K_{\perp\parallel}^2 +  k^2)  \psi 
+ 2 K_{\perp\parallel}{\dot \phi} 
&=& 0\,, \\
{\ddot\phi} - K_{\perp\parallel}^2\phi 
- 2 K_{\perp\parallel} \dot\psi  &=& 0\,,
\\
{\ddot\phi}_2 =0\,.
\end{eqnarray}

Let 

\begin{eqnarray}
\psi &=& \psi_0 e^{-i\omega \tau}\,, \nonumber \\
\phi &=& \phi_0 e^{-i\omega \tau}\,. \nonumber
\end{eqnarray}
We then have

\begin{eqnarray}
\left(\omega^2 + K_{\perp\parallel}^2  + 
k^2 \right)  \psi_0 
+ 2i\omega  K_{\perp\parallel}\phi_{0}&=& 0\,,\nonumber\\
\left(\omega^2 +  K_{\perp\parallel}^2 \right)\phi_{0} 
- 2i\omega K_{\perp\parallel} \psi_0  &=& 0\,.
\end{eqnarray}
A non-trivial solution exists when

\begin{equation}
\omega^4 +  \left[ k^2 - 
2 K_{\perp\parallel}^2 \right] \omega^2 + 
\left[ k^2 + 
 K_{\perp\parallel}^2 \right]
K_{\perp||}^2 = 0\,.
\end{equation}
When we substitute the values (\ref{eq:kmx}) and (\ref{eq:gc}), we find that
all four eigenfrequencies are complex.
This implies 
that no
analogue of the breathing modes exists in the 
DNG spectrum. 

The linearized  theory is unusual in two related respects.
There is no feedback on $\partial m$ from perturbations in the 
bulk. The boundary equations of motion can be solved without reference to 
the bulk motion. In particular, the two boundaries 
behave independently at this order.
The physical reason for this is the weak nature of the coupling 
between the string and the end particles. The force which 
binds them  is constant, not harmonic.

\section{Conclusions}

In this paper, we have derived the complete linearized 
equations of motion for a relativistic DNG membrane
with DNG edges. One direct physical application 
of these equations is to examine the 
stability of the QCD string with massive quarks at its ends.
For rigidly rotating configurations,
we calculate the normal modes 
and we find that, in general, they are complex. Whether this 
implies a genuine instability in the system remains to be demonstrated.
The necessary classical ingredient in the determination of the
effect of quantum fluctuations in a saddle point evaluation
of the path integral is the calculation of the 
second order variation of the action. 
This variation can be read off straightforwardly 
from the expressions derived in this paper. 

Finally, we mention that relativistic
extended objects with edges provide a simplified 
well defined setting in which
to examine the relationship between 
bulk and boundary degrees of freedom
that is thought to be relevant in various 
contexts in gravitational
physics, most notably that of black holes.

\section*{Acknowledgments}

We have benefitted from conversations with Brandon Carter, Xavier Martin
and Alexander Vilenkin.
We gratefully acknowledge support from CONACyT grant no.
211085-5-0118PE. JG would like to thank Prof. J. Lewis 
of the Dublin Institute for Advanced Studies for
hospitality during his stay in Dublin.

\section*{Appendix}

In this appendix, we give a brief list of the equations corresponding 
to the case of an arbitrary background spacetime. 
We denote with $R^\mu{}_{\nu \rho \sigma}$ the
Riemann curvature tensor of the background spacetime.

The extrinsic curvature of the edge as embedded in
spacetime transforms now as
\begin{equation}
\hat{D}_{\delta\bar X_\perp} L_{AB}{}^I
= -\hat{\cal D}_A \hat{\cal D}_B \phi^I +  L_{AC}{}^I 
L^C{}_{B\,J}  \phi^J - R_{\mu \nu \sigma \rho}
f^\mu{}_A f^\sigma{}_B m^{\nu \, I} m^\rho{}_J \phi^J \,.
\end{equation}

The additional term involving the background curvature
propagates in the equations corresponding to the projections 
as

\begin{eqnarray}
\hat D_{\delta\bar X_\perp} k_{AB}
=   &-& {\cal D}_A {\cal D}_B \psi + \left[ K^i_A K_{i\,B} 
+ k_{AC} k^C{}_B -
R_{\mu \nu \sigma \rho}
f^\mu{}_A f^\sigma{}_B \eta^{\nu} \eta^\rho \right]  \psi \nonumber \\
&-& 2 K^i{}_{(A}\widetilde{\cal D}_{B)}\phi_i
- (\widetilde{\cal D}_A K^i_B ) \, \phi_i 
+  k_{AC}  K^C_{B\,i} \phi^i
- R_{\mu \nu \sigma \rho}
f^\mu{}_A f^\sigma{}_B \eta^{\nu} n^\rho{}_i \phi^i\,,
\end{eqnarray}
and

\begin{eqnarray}
\hat{D}_{\delta\bar X_\perp} K_{AB}{}^i
= &-&\widetilde{\cal D}_A \widetilde{\cal D}_B \phi^i +
\left[ K_{AC}{}^i K^C{}_{B\,j} + K_A{}^i K_{B\,j} -
R_{\mu \nu \sigma \rho}
f^\mu{}_A f^\sigma{}_B n^{\nu \i } n^\rho{}_j \right]
\phi^j
\nonumber\\
&+&  K_{AC}{}^i k^C{}_{B} \psi + (\widetilde{\cal D}_A K_B{}^i) \psi +
2 K_{(A}{}^i {\cal D}_{B)}\psi -
R_{\mu \nu \sigma \rho}
f^\mu{}_A f^\sigma{}_B n^{\nu \, i} \eta^\rho \psi\,.
\end{eqnarray}

For the traces over the edge indices, we obtain:

\begin{eqnarray}
\hat{D}_{\delta\bar X_\perp} k = 
= &-& {\cal D}_A{\cal D}^A \psi 
+ \left[K_{A\,i} K^{A\,i} - k_{AB}k^{AB}  
  R_{\mu \nu \sigma \rho}
f^\mu{}_A f^{\sigma \, A} \eta^{\nu} \eta^\rho    \right]\psi\nonumber\\
&-&  
2 K_{A}{}^i (\widetilde{\cal D}^A\phi_i )
- (\widetilde{\cal D}_A K^{A\,i} ) \, \phi_i
- k_{AB} K^{AB}{}_i  \phi^i \nonumber \\
&-& R_{\mu \nu \sigma \rho}
f^\mu{}_A f^{\sigma \,B} \eta^{\nu} n^\rho{}_i \phi^i\,,
\label{eq:dck}
\end{eqnarray}
and

\begin{eqnarray}
\hat{D}_{\delta\bar X_\perp} \left( h^{AB} K_{AB}{}^i \right)
= &-&\widetilde{\cal D}_A \widetilde{\cal D}^A \phi^i -
\left[ K_{AB}{}^i K^{AB\,j} -  K_A{}^i K^A{}_j 
 - R_{\mu \nu \sigma \rho}
f^\mu{}_A f^{\sigma \, B} n^{\nu \i } n^\rho{}_j \right] \phi^j \nonumber\\
&-&  K_{AB}{}^i k^{AB}\psi  - (\widetilde{\cal D}_A K^{A\,i} ) \psi 
+   2 K_{A}{}^i {\cal D}^A\psi \nonumber \\
&-& R_{\mu \nu \sigma \rho}
f^\mu{}_A f^{\sigma \,B} n^{\nu \, i} \eta^\rho \psi\,.
\label{eq:dcK}
\end{eqnarray}

The linearized equations of motion for the bulk 
are given by
\cite{GUV,LF,CAR,CG1}:

\begin{equation}
\tilde\Delta \Phi^i + 
[ K_{ab}{}^i K^{ab}{}_j  
-
R_{\mu \nu \sigma \rho}
e^\mu{}_a e^{\sigma \, a}  n^{\nu \, i} n^\rho{}_j ]  \Phi^j 
= 0 \,,
\end{equation}

and, finally, the linearized equations of motion for the
edges are obtained by simply setting to zero the
right hand sides of Eqs. (\ref{eq:dck}), (\ref{eq:dcK}).

\end{document}